\documentclass[preprint,prb, amsmath,amssymb,aps,superscriptaddress,nobibnotes]{revtex4-1}
\usepackage{CJK}
\usepackage{graphicx}
\usepackage{color}

\usepackage{dcolumn} 
\usepackage{bm}     
\usepackage{amsfonts}
\usepackage[bookmarks=false,colorlinks,citecolor=red]{hyperref}
\usepackage{amsmath}

\usepackage[version=4]{mhchem}
\usepackage{siunitx}

\newcommand{\bfl}{\begin{flushleft}}
\newcommand{\efl}{\end{flushleft}}
\include{MyCommand}

\newif\ifshowcomments\showcommentstrue

\begin{document}

\title{
Superconductivity in a quintuple-layer square-planar nickelate
}

\date{\today}
\author{Grace A. Pan}
\affiliation{Department of Physics, Harvard University, Cambridge, MA, USA}
\author{Dan Ferenc Segedin}
\affiliation{Department of Physics, Harvard University, Cambridge, MA, USA}
\author{Harrison LaBollita}
\affiliation{Department of Physics, Arizona State University, Tempe, AZ, USA}
\author{Qi Song}
\affiliation{Department of Physics, Harvard University, Cambridge, MA, USA}
\author{Emilian M. Nica}
\affiliation{Department of Physics, Arizona State University, Tempe, AZ, USA}
\author{Berit H. Goodge}
\affiliation{School of Applied and Engineering Physics, Cornell University, Ithaca, NY, USA}
\affiliation{Kavli Institute at Cornell for Nanoscale Science, Cornell University, Ithaca, NY, USA}
\author{Andrew T. Pierce}
\affiliation{Department of Physics, Harvard University, Cambridge, MA, USA}
\author{Spencer Doyle}
\affiliation{Department of Physics, Harvard University, Cambridge, MA, USA}
\author{Steve Novakov}
\affiliation{Department of Physics, University of Michigan, Ann Arbor, MI, USA}
\author{Denisse C\'{o}rdova Carrizales}
\affiliation{Department of Physics, Harvard University, Cambridge, MA, USA}
\author{Alpha T. N'Diaye}
\affiliation{Advanced Light Source, Lawrence Berkeley National Laboratory, Berkeley, CA, USA}
\author{Padraic Shafer}
\affiliation{Advanced Light Source, Lawrence Berkeley National Laboratory, Berkeley, CA, USA}
\author{Hanjong Paik}
\affiliation{Platform for the Accelerated Realization, Analysis, and Discovery of Interface Materials (PARADIM), Cornell University, Ithaca, NY, USA}
\author{John T. Heron}
\affiliation{Department of Materials Science and Engineering, University of Michigan, Ann Arbor, MI, USA}
\author{Jarad A. Mason}
\affiliation{Department of Chemistry, Harvard University, Cambridge, MA, USA}
\author{Amir Yacoby}
\affiliation{Department of Physics, Harvard University, Cambridge, MA, USA}
\author{Lena F. Kourkoutis}
\affiliation{School of Applied and Engineering Physics, Cornell University, Ithaca, NY, USA}
\affiliation{Kavli Institute at Cornell for Nanoscale Science, Cornell University, Ithaca, NY, USA}
\author{Onur Erten}
\affiliation{Department of Physics, Arizona State University, Tempe, AZ, USA}
\author{Charles M. Brooks}
\affiliation{Department of Physics, Harvard University, Cambridge, MA, USA}
\author{Antia S. Botana}
\thanks{Correspondence should be addressed to: \href{mailto:antia.botana@asu.edu}{antia.botana@asu.edu}; \href{mailto:mundy@fas.harvard.edu}{mundy@fas.harvard.edu}}
\affiliation{Department of Physics, Arizona State University, Tempe, AZ, USA}
\author{Julia A. Mundy}
\thanks{Correspondence should be addressed to: \href{mailto:antia.botana@asu.edu}{antia.botana@asu.edu}; \href{mailto:mundy@fas.harvard.edu}{mundy@fas.harvard.edu}}
\affiliation{Department of Physics, Harvard University, Cambridge, MA, USA}

\maketitle

\noindent \textbf{Since the discovery of high-temperature superconductivity in the copper oxide materials \cite{highTcCuprate}, there have been sustained efforts to both understand the origins of this phase and discover new cuprate-like superconducting materials \cite{review_cuprates}. One prime materials platform has been the rare-earth nickelates and indeed superconductivity was recently discovered in the doped compound Nd$_{0.8}$Sr$_{0.2}$NiO$_2$ \cite{Li2019}.  Undoped NdNiO$_2$ belongs to a series of layered square-planar nickelates with chemical formula Nd$_{n+1}$Ni$_n$O$_{2n+2}$ and is known as the `infinite-layer' ($n = \infty$) nickelate. Here, we report the synthesis of the quintuple-layer ($n = 5$) member of this series, Nd$_6$Ni$_5$O$_{12}$, in which optimal cuprate-like electron filling ($d^{8.8}$) is achieved without chemical doping.  We observe a superconducting transition beginning at $\sim$13 K.  Electronic structure calculations, in tandem with magnetoresistive and spectroscopic measurements, suggest that Nd$_6$Ni$_5$O$_{12}$ interpolates between cuprate-like and infinite-layer nickelate-like behavior.  In engineering a distinct superconducting nickelate, we identify the square-planar nickelates as a new family of superconductors which can be tuned via both doping and dimensionality.}

Attempts to construct materials that are isoelectronic and isostructural to the cuprates as candidate superconducting systems have spanned more than three decades \cite{review_cuprates}. In this context, nickelate materials have been of enduring interest due to the proximity of nickel and copper on the periodic table \cite{anisimov}; furthermore, Ni$^{1+}$ has the same $d^9$ electron count as Cu$^{2+}$. With continued understanding of the hallmarks of cuprate superconductivity – including for example a prominent orbital polarisation, strong transition metal-oxygen hybridisation, or a quasi-two-dimensional square lattice – a general attention to nickelates has been refined to specific proposals for LaNiO$_3$/La$M$O$_3$ superlattices ($M = $ trivalent cation) \cite{chaloupka2008orbital} and La$_{2-x}$Sr$_x$NiO$_4$ \cite{emery}. To date, superconductivity has not been realised in any of these originally proposed materials.  Most recently however, Sr-doped  NdNiO$_2$, the so-called `infinite-layer' nickelate, has been shown to be superconducting \cite{Li2019}, with a dome-like doping dependence for $T_c$ centred around $d^{8.8}$ nickel filling \cite{Li2020_SC_dome,zeng2020phase}. 

Following this observation of superconductivity in Sr-doped NdNiO$_2$, there has been considerable interest in determining how `cuprate-like' these infinite-layer nickelates are.  For example, while the cuprates are often described by a single-band Hubbard model, electronic structure descriptions of the infinite-layer nickelates invoke multi-band models including hybridisation with the rare-earth 5$d$ bands \cite{pickett, prx,  deveraux}. Infinite-layer nickelates also seem to exhibit a much larger charge-transfer energy than their cuprate counterparts \cite{deveraux, prx, goodge2020}. Experimentally, the synthesis of these compounds remains challenging \cite{hwang_synthesis} and to date there are only three independent reports of superconductivity \cite{Li2019,zeng2020phase,tunneling}, all in the chemically doped infinite-layer compound.  These extended difficulties in reproducing nickelate superconductivity have generated a flurry of questions involving the synthetic challenges, the substrate-film interface \cite{Li2020absence,BiXiaWang2020}, and most importantly, whether there exists an entire family of nickelate superconductors beyond compounds accessible through isovalent cation substitution.

\begin{figure}
    \centering
    \includegraphics[width = \columnwidth]{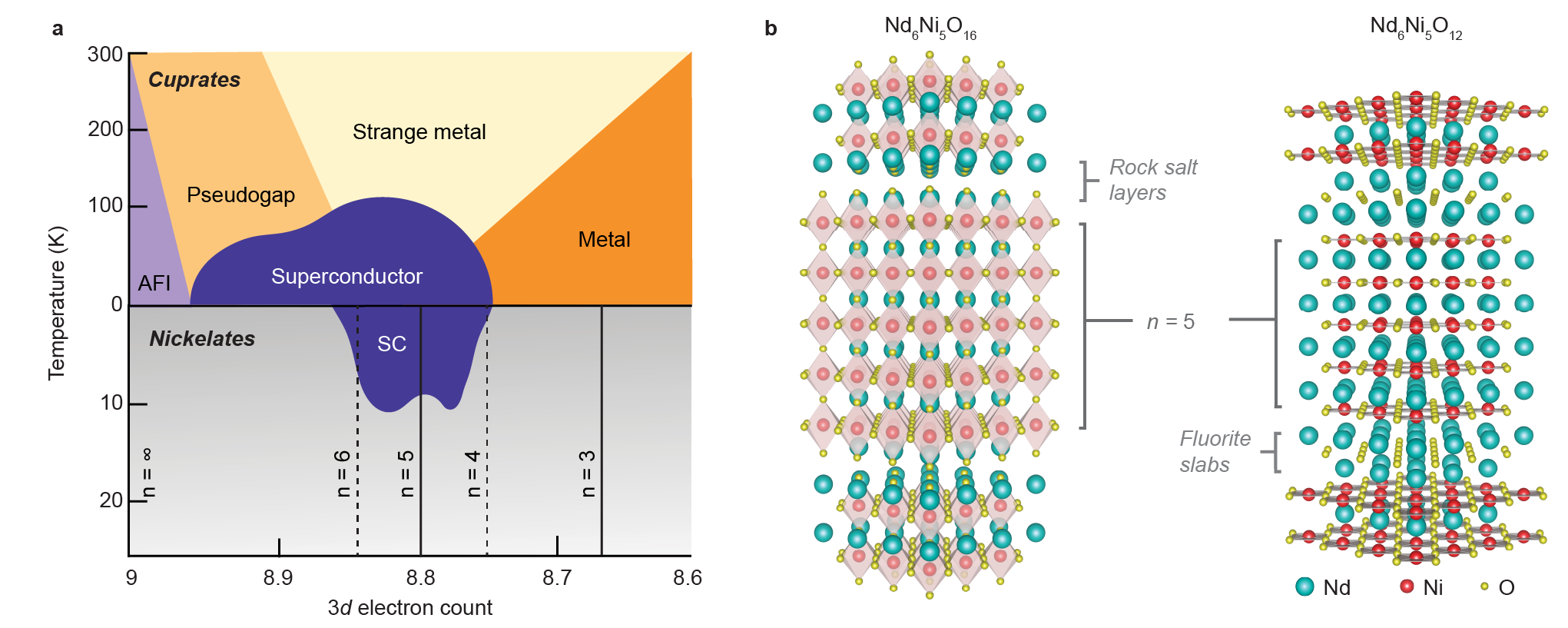}
    \caption{\textbf{Electronic phase diagram and structural description of the layered nickelates. a,} Schematic phase diagram for the electronic phases of the cuprates (top) and nickelates (bottom). The superconducting dome for the cuprates is adapted from ref. \onlinecite{review_cuprates}; for the nickelates, from the Sr-doped NdNiO$_2$ \cite{Li2020_SC_dome,zeng2020phase}. The square-planar $R_{n+1}$Ni$_n$O$_{2n+2}$ phases are identified on the phase diagram by their formal 3$d$ electron count. AFI, antiferromagnetic insulator; SC, superconductor. \textbf{b,} Crystal structures of the quintuple-layer nickelates in the Nd$_6$Ni$_5$O$_{16}$ Ruddlesden-Popper phase (left) and Nd$_6$Ni$_5$O$_{12}$ reduced square-planar phase (right), depicted at the same scale. Note the change in oxygen coordination upon reduction: the octahedral coordination in the five layer structure converts to the square-planar coordination, and the rock salt layers separating every five nickel layers transform into fluorite blocking slabs.  Nd, Ni and O in turquoise, red and yellow respectively.}
    \label{fig:structure}
\end{figure}

Here, we note that NdNiO$_2$ is the end, or `infinite' ($n = \infty$), member compound of the broader layered $R_{n+1}$Ni$_n$O$_{2n+2}$ ($R$ = trivalent rare-earth cation, $n > 1$) series.  This family of layered materials possesses  quasi-two-dimensional NiO$_2$ planes in a square-planar coordination with rare-earth fluorite blocking slabs interleaved every $n$ nickel layers.  These square-planar compounds are obtained via oxygen deintercalation from the corresponding parent perovskite $R$NiO$_3$ ($n = \infty$) and Ruddlesden-Popper  $R_{n+1}$Ni$_n$O$_{3n+1}$ ($n \neq \infty$) phases (Fig. \ref{fig:structure}b). As shown in Fig. \ref{fig:structure}a, the $R_{n+1}$Ni$_n$O$_{2n+2}$  series can be mapped onto the cuprate phase diagram in terms of the nickel 3$d$-electron count using formal valence counting rules (Supplementary Note 1), where the end member NdNiO$_2$ has $d^9$ (Ni$^{1+}$) filling.  The trilayer $R_4$Ni$_3$O$_{8}$ ($n = 3$, $R$ = La, Pr) nickelates have been reported as close cuprate analogues even though superconductivity has not been observed \cite{nat_phys}. Both La$_4$Ni$_3$O$_{8}$ and Pr$_4$Ni$_3$O$_{8}$ exhibit strong orbital polarisation \cite{nat_phys} with a large cuprate-like magnetic superexchange \cite{dean}, strengthening the connection to the cuprate phase diagram. These trilayer nickelates have an average $d^{8.67}$ (Ni$^{1.33+}$) filling across the nickel planes (note that in multilayer cuprates, NMR experiments have shown that there is a layer-resolved doping profile \cite{multilayer_cuprates}), lying in the ``overdoped" regime of the cuprate phase diagram. Meanwhile, the `quintuple-layer' compound $R_6$Ni$_5$O$_{12}$ ($n$ = 5) with average $d^{8.8}$ (Ni$^{1.2+}$) filling falls directly at optimal doping to align with the cuprate phase diagram and is thus an appealing target to search for further nickelate superconductivity.

We synthesize the square-planar layered compound Nd$_6$Ni$_5$O$_{12}$ ($n$ = 5) with optimal nickel $d^{8.8}$ filling from its Ruddlesden-Popper parent phase Nd$_6$Ni$_5$O$_{16}$.  We additionally synthesize the ``overdoped" layered compound Nd$_4$Ni$_3$O$_8$ ($n$ = 3) with $d^{8.67}$ filling, whose Ruddlesden-Popper parent phase Nd$_4$Ni$_3$O$_{10}$ has been well-characterised in the bulk, in order to benchmark our studies \cite{GREENBLATT1997174,li2021contrasting}. First, we employ reactive-oxide molecular beam epitaxy (MBE) to stabilise epitaxial thin films of the oxidised Ruddlesden-Popper Nd$_{n+1}$Ni$_n$O$_{3n+1}$ compounds on (110)-orientated NdGaO$_3$ substrates (Methods, Supplementary Note 2).  This is followed by an $ex$ $situ$ CaH$_2$-assisted topotactic reduction to deintercalate oxygen and convert the films into the square-planar (Nd$_{n+1}$Ni$_n$O$_{2n+2}$) phase (Methods), as has been performed in the infinite-layer compounds \cite{Li2019,hwang_synthesis}.  

\begin{figure}
    \centering
    \includegraphics[width = 1\columnwidth]{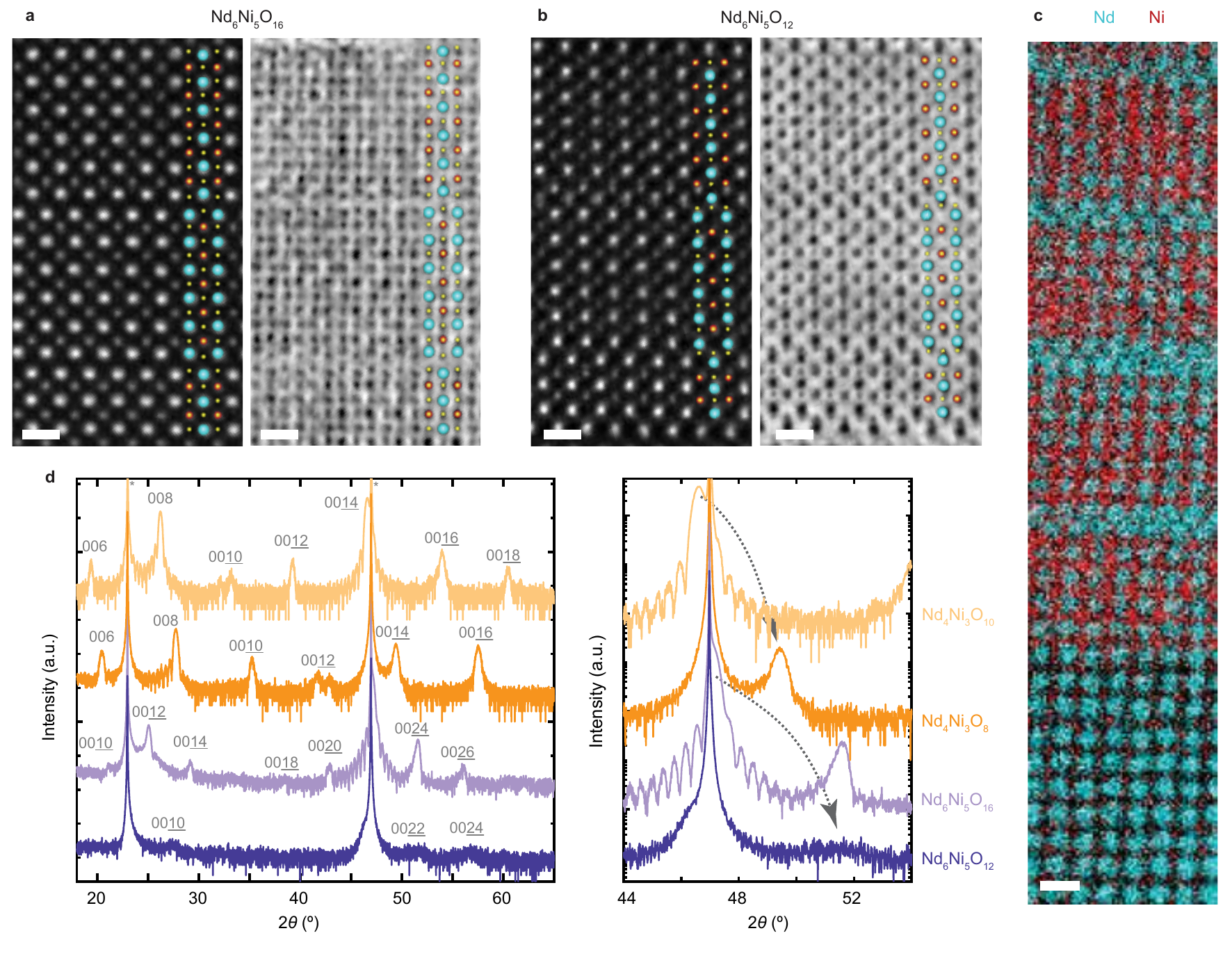}
    \caption{\textbf{Structural characterisation of the layered nickelates. a, b} High angle annular dark-field (left) and annular bright field (right) scanning transmission electron microscopy images of the Nd$_6$Ni$_5$O$_{16}$ \textbf{(a)} and Nd$_6$Ni$_5$O$_{12}$ \textbf{(b)} compounds. Nd, Ni and O are represented by turquoise, red and yellow dots, respectively. Scale bars, 5 \AA. \textbf{c,} Electron energy loss spectroscopy map of the Nd$_6$Ni$_5$O$_{12}$ compound near the substrate-film interface. Scale bar, 5 \AA. \textbf{d,} X-ray diffraction spectra of the $n$ = 3 and $n$ = 5 compounds over the full scan range (left). Substrate peaks are labelled with asterisks and curves are vertically offset for clarity.  A zoomed-in region (right) illustrates how the main superlattice peaks ((00\underline{14}) for $n = 3$ and (00\underline{22}) for $n = 5$) shift upon reduction, as indicated by the grey arrows.}
    \label{fig:mainstructural}
\end{figure}

Cross-sectional scanning transmission electron microscopy (STEM) images of the $n = 5$ parent Ruddlesden-Popper phase Nd$_6$Ni$_5$O$_{16}$ and the reduced compound Nd$_6$Ni$_5$O$_{12}$ are shown in Figs. \ref{fig:mainstructural}a-c. Atomic-resolution high angle annular dark field (HAADF) and annular bright field (ABF) images of well-ordered regions confirm both the shortening of the $c$-axis lattice constant and the removal of apical oxygen in the neodymium planes from the Nd$_6$Ni$_5$O$_{16}$ (Fig. \ref{fig:mainstructural}a) to the Nd$_6$Ni$_5$O$_{12}$ phase (Fig. \ref{fig:mainstructural}b). As shown in Fig. \ref{fig:mainstructural}c, elemental mapping by electron energy loss spectroscopy (EELS) illustrates the quintuple-layer structure with the neodymium fluorite layers of the Nd$_6$Ni$_5$O$_{12}$ film. Similar to  superconducting infinite-layer nickelates \cite{hwang_synthesis,goodge2020}, these samples also show regions of secondary or defect inclusions in addition to the layered nickelate phase (Supplementary Note 3, Figs. S4-S7). The structural coherence at the bulk scale of both $n = 3$  and $n = 5$ Ruddlesden-Popper and square-planar phases was probed with x-ray diffraction (XRD).  XRD scans in Fig. \ref{fig:mainstructural}d display the presence of superlattice peaks demonstrating ordering of the Ruddlesden-Popper structure.  Upon reduction, we observe a rightward shift of the superlattice peaks in the diffraction spectra corresponding to a compression of the out-of-plane lattice constants from $c$ = 42.5 \AA{} to $c$ = 38.8 \AA{} for Nd$_6$Ni$_5$O$_{12}$ and $c$ = 27.4 \AA{} to $c$ = 25.4 \AA{} for Nd$_4$Ni$_3$O$_{8}$ consistent with the STEM imaging.  Electronic characterisation via EELS (Supplementary Note 3, Fig. S8) and x-ray absorption spectroscopy (XAS) (Supplementary Note 5, Fig. S11) further confirms reduction to the square-planar phase. The intense pre-peak at the O K-edge in Nd$_6$Ni$_5$O$_{16}$ associated with oxygen 2$p$ orbitals hybridised with (formal) nickel 3$d^{7+\delta}$ states is altered upon reduction \cite{abbate2002electronic}.  A shoulder with weaker spectral weight emerges at higher energies in Nd$_6$Ni$_5$O$_{12}$, consistent with the hybridised nickel 3$d^{9-\delta}$ - oxygen 2$p$ states.  This has been seen in bulk compounds \cite{nat_phys,dean}, as well as in infinite-layer compounds which achieve similar nickel filling through chemical doping \cite{goodge2020}.

\begin{figure}
    \centering
    \includegraphics[width = 1.\columnwidth]{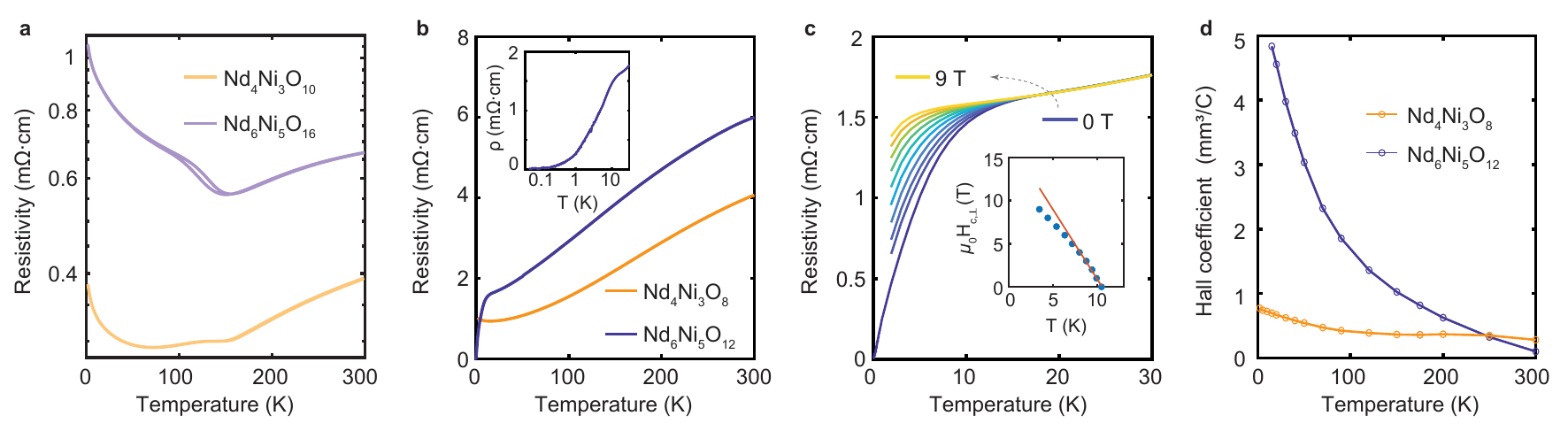}
    \caption{\textbf{Transport properties of the layered nickelate thin films. a,} Resistivity $\rho$(T) for the unreduced Nd$_4$Ni$_3$O$_{10}$ ($n$ = 3) and Nd$_6$Ni$_5$O$_{16}$ ($n$ = 5) Ruddlesden-Popper compounds.  \textbf{b,} Resistive transitions of the layered square-planar Nd$_4$Ni$_3$O$_8$ ($n$ = 3) and Nd$_6$Ni$_5$O$_{12}$ ($n$ = 5) compounds into the insulating upturn and superconducting states, respectively.  Inset is a close up of the superconducting transition of the $n$ = 5 compound from 50 mK to 30 K with temperature plotted on a log scale.  \textbf{c,} Resistivity $\rho(T)$ of the Nd$_6$Ni$_5$O$_{12}$ ($n$ = 5) below $T = 30$ K for $\mu_0 H_{c,\perp}$ = 0 - 9 T in increments of 1 T.  The zero-field resistivity goes down to 50 mK; all else to 1.8 K.  Inset shows the relationship between $\mu_0 H_{c,\perp}$ and $T_c$, with a linear scaling and fit near $T_c$.  \textbf{d,} Temperature-dependent Hall coefficients for the reduced Nd$_4$Ni$_3$O$_8$ and Nd$_6$Ni$_5$O$_{12}$ compounds.}
    \label{fig:maintransport}
\end{figure}

The electronic properties of the layered nickelate thin-films are shown in Fig. \ref{fig:maintransport}.  Figure \ref{fig:maintransport}a shows the temperature-dependent resistivities of the as-synthesised Ruddlesden-Popper compounds before reduction.  The Nd$_4$Ni$_3$O$_{10}$ film exhibits a resistivity kink at $\sim$155 K, likely due to charge density wave formation \cite{GREENBLATT1997174,li2021contrasting}.  Our Nd$_6$Ni$_5$O$_{16}$ Ruddlesden-Popper films all display a metal-to-insulator transition with a weak hysteresis reminiscent of the first-order transition of the perovskite NdNiO$_3$.  Upon reduction, Nd$_4$Ni$_3$O$_{8}$ exhibits metallic behavior similar to that observed in the infinite-layer NdNiO$_2$ \cite{Li2019} and trilayer Pr$_4$Ni$_3$O$_8$ \cite{nat_phys}; we note that the ground state behavior of bulk Nd$_4$Ni$_3$O$_{8}$ has not yet been well-established \cite{li2021contrasting,miyatake2020chemical}.  Meanwhile, Nd$_6$Ni$_5$O$_{12}$ becomes superconducting (Fig \ref{fig:maintransport}b).  Superconductivity begins to onset as high as $\sim$15 K with the point of maximum curvature at $\sim$13 K.  The resistivity reaches half the normal state value at $\sim$4.5 K and hits zero Ohms by $\sim$ 100 mK as shown in the inset.  The suppression of superconductivity by a magnetic field up to 9 T in the $c$-axis perpendicular to the plane of the sample is presented in Fig. \ref{fig:maintransport}c (see also Supplementary Note 4, Fig. S9).  The normal state resistivity is approximately linear in temperature immediately before approaching the superconducting transition and responds minimally to an out-of-plane magnetic field.  Due to the breadth of the transition, we use as a proxy for the upper critical field $H_{c,\perp}$ the strength of the field where the resistivity reaches 90\% of the normal state value where superconducting correlations are present \cite{wang2020isotropic}.  Linear in the Ginzburg-Landau model (inset, Fig. \ref{fig:maintransport}c; Methods) around $T_c$, this allows us to estimate an in-plane correlation length of $\xi(T=0)_{ab}$ = 4.4 $\pm$ 0.2 nm, comparable to that of the doped infinite-layer compound \cite{Li2019,wang2020isotropic}.

Particularly noteworthy in the context of the infinite-layer nickelates is the temperature dependence of the Hall coefficient (Fig. \ref{fig:maintransport}d).  The doped infinite-layer nickelates possess Hall coefficients that, while sensitive to factors such as doping levels \cite{zeng2020phase,Li2020_SC_dome,Osada2020}, thickness \cite{zeng2021observation}, and cation composition \cite{Osada2020,osada2021nickelate,zeng2021superconductivity}, ultimately remain negative at all temperatures or exhibit a zero-crossing.  This evinces a two-band picture with substantive electron- and hole-pocket contributions \cite{Li2019,deveraux,prx}.  Meanwhile, both the superconducting Nd$_6$Ni$_5$O$_{12}$ and non-superconducting Nd$_4$Ni$_3$O$_8$ layered nickelates have Hall coefficients that remain positive at all temperatures, with a semiconductor-like temperature dependence reminiscent of the under- and optimally-doped layered cuprates \cite{takagi1989superconductor}.  For reference, we note that the Hall coefficients of oxidised Ruddlesden-Popper thin films behave as expected (Fig. S10): the sudden jump in the Hall coefficient ($R_\text{H}$) below $\sim$155 K for Nd$_4$Ni$_3$O$_{10}$ likely reflects a Fermi surface reconstruction due to a charge density wave instability and the abrupt loss of carriers as has been observed in the bulk \cite{li2021contrasting}. 

\begin{figure}
    \centering
    \includegraphics[width = 0.5\columnwidth]{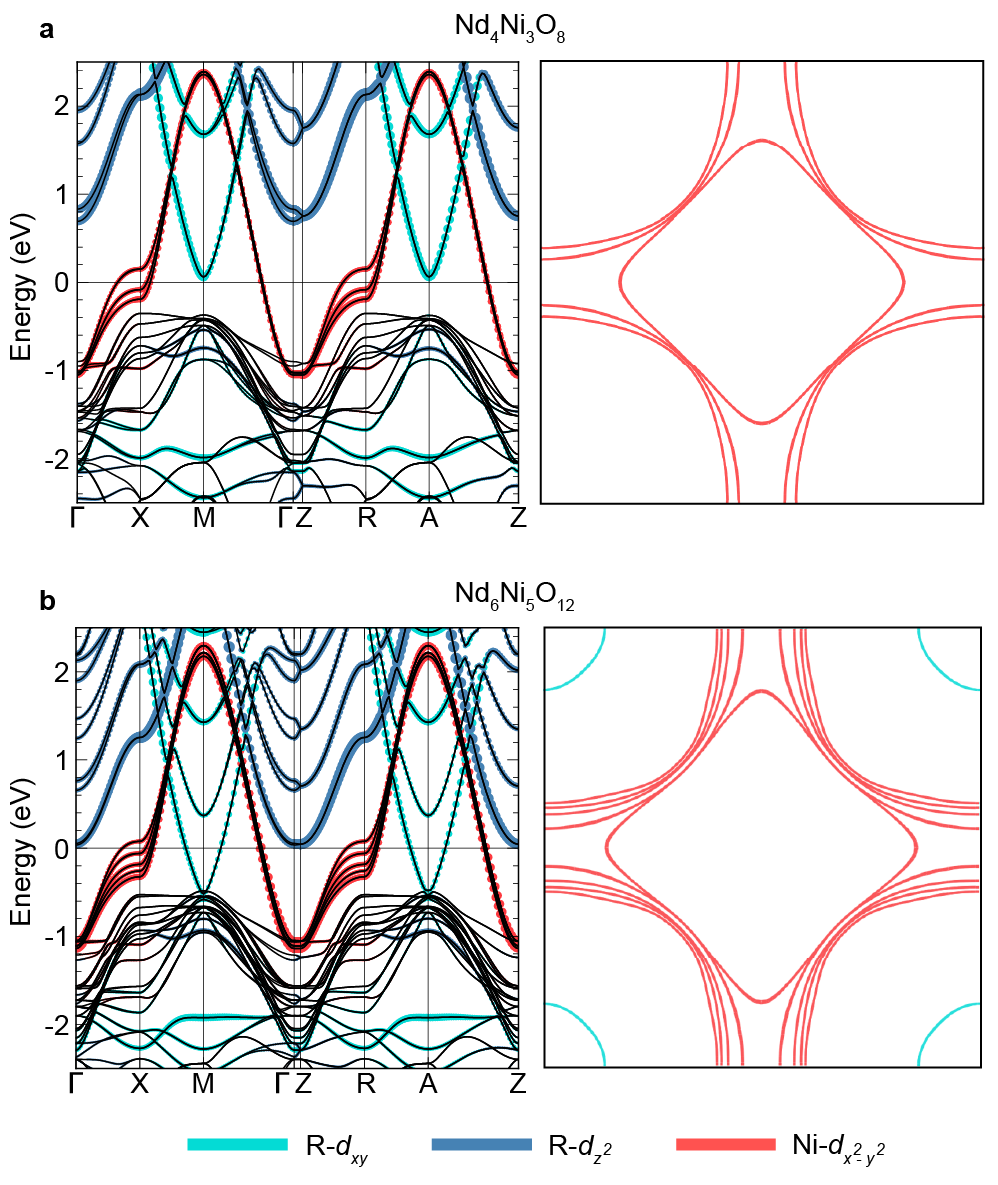}
    \caption{{\bf Electronic structure description of the layered nickelates. a, b} Layered Nd$_4$Ni$_3$O$_8$ ($n=3$) \textbf{(a)} and Nd$_6$Ni$_5$O$_{12}$ ($n=5$) \textbf{(b)} paramagnetic band structures and corresponding Fermi surfaces with Nd-$d_{z^{2}}$, Nd-$d_{xy}$, and Ni-$d_{x^{2}-y^{2}}$ orbital-site content (`fatbands') highlighted.}
    \label{fig:ES}
\end{figure}

With this observation, it becomes relevant to compare the predicted electronic structure of Nd$_6$Ni$_5$O$_{12}$ ($n= 5$) square-planar layered nickelates with their infinite-layer NdNiO$_2$ ($n= \infty$) and trilayer Nd$_4$Ni$_3$O$_8$ ($n=3$) counterparts (see Fig. \ref{fig:ES}). In NdNiO$_2$ ($d^9$ filling), much attention has been paid to its multiband character: in the paramagnetic state, in addition to the Ni-$d_{x^2-y^2}$ bands, Nd-$d$ states also cross the Fermi level. The latter lead to the appearance of electron pockets
at the $\Gamma$ and A points which display mainly Nd-$d_{z^2}$ and Nd-$d_{xy}$ character, respectively. These electron pockets give rise to self-doping of the large hole-like Ni-$d_{x^2-y^2}$ Fermi surface \cite{pickett, prx, Thomale_PRB2020}. Notably, the electronic structure is three-dimensional-like, with a large $c$-axis dispersion. In contrast, in Nd$_4$Ni$_3$O$_8$ (at $d^{8.67}$ filling), a single band per nickel crosses the Fermi level and there is no Nd-$d$ involvement; this leads to a much more cuprate-like scenario \cite{karp_prb}. 
The electronic structure of Nd$_6$Ni$_5$O$_{12}$ seems to interpolate between these two situations: a single $d_{x^{2} -y^{2}}$ band per nickel crosses the Fermi level but Nd-$5d$ bands start playing a role, even though the electron pockets related to Nd-$5d$ states are significantly smaller than those in the infinite-layer material \cite{labollita2021elec}. These features of the electronic structure suggest that Nd-Ni 5$d$-3$d$ hybridization is less important in Nd$_6$Ni$_5$O$_{12}$ than in the infinite-layer compounds. Aside from the appearance of the neodymium-derived pockets at the zone corners, the Fermi surface of Nd$_6$Ni$_5$O$_{12}$ is analogous to that of multilayer cuprates with one electron-like and four hole-like $d_{x^2-y^2}$ Fermi surface sheets \cite{sakakibara}. Importantly, the Fermi surface of the quintuple layer nickelate is much more two-dimensional-like compared to the infinite-layer nickelate compound as the presence of the fluorite blocking layer cuts the $c$-axis dispersion. 
From our description of the electronic structure, we calculate low-temperature Hall coefficients ($R_\text{H}$) using semi-classical Boltzmann transport theory (Supplementary Note 6). In the $T$ = 0 limit, we find $R_{\text{H}} \sim 1$ mm$^{3}/$C and $5$ mm$^{3}/$C for Nd$_{4}$Ni$_{3}$O$_{8}$ and Nd$_{6}$Ni$_{5}$O$_{12}$, respectively. These values are consistent with the experimental low-temperature $R_\text{H}$ presented above and suggest that even at identical $d^{8.8}$ filling levels, the layered Nd$_{6}$Ni$_{5}$O$_{12}$ compound has a weaker involvement of the Nd-$d$ bands and is more cuprate-like than its Sr-doped NdNiO$_2$ counterpart.

In the cuprates, a large $p$-$d$ hybridisation is an essential ingredient in mediating high-temperature superconductivity \cite{zhangRice}. A quantitative measurement of this is the charge transfer energy $\Delta$, which is inversely proportional to the hybridisation. In the context of infinite-layer superconducting nickelates, the larger charge-transfer energy  ($\Delta_{112} \sim4.4$ eV) \cite{prx,goodge2020} compared to cuprates has been intensively analysed. From  electronic structure calculations, $\Delta$ can be obtained as the difference in energy between the Ni-$d_{x^{2} - y^{2}}$ and the O-$p\sigma$ on-site energies derived from maximally-localised Wannier functions (Supplementary Note 6).  The charge-transfer energy for Nd$_4$Ni$_3$O$_8$ is $\Delta_{438} \sim3.5$ eV, consistent with previous work \cite{nica2020}, and for Nd$_6$Ni$_5$O$_{12}$ is $\Delta_{6512} \sim 4.0$ eV.  These are  significantly smaller values than those obtained for the infinite-layer compounds. We can qualitatively assess this calculation using x-ray absorption spectroscopy (XAS) on our synthesised films. Spectra at the O K-edge for the Nd$_6$Ni$_5$O$_{12}$ and Nd$_4$Ni$_3$O$_{8}$ compounds exhibit a pre-peak feature whose intensities increase with diminishing dimensionality (Fig. S11). Since the pre-peak corresponds to excitations from the O-1$s$ states to hybridised Ni-O 2$p$-3$d$ states, comparisons of pre-peak intensities can be used as a proxy to estimate the relative size of the charge transfer gap or transition-metal-oxygen hybridisation \cite{suntivich2014estimating}. Meanwhile, no pre-peak has been observed in undoped infinite-layer nickelates \cite{deveraux}. Hence, our calculations and XAS jointly suggest a slightly reduced charge transfer energy and enhanced covalency in the layered nickelates compared to the infinite-layer compounds.

We have synthesised a quintuple-layer square-planar nickelate that demonstrates superconductivity with a $T_c$ of $\sim$13 K.  Our square-planar trilayer and quintuple-layer compounds possess similarities to the doped infinite-layer nickelates but notably have positive Hall coefficients up to room temperature indicative of a more single-band-like picture.  This, in tandem with our theoretical description, suggests that the quintuple-layer Nd$_6$Ni$_5$O$_{12}$ compound possesses an electronic structure that is qualitatively intermediate between that of the cuprates and infinite-layer nickelates, and tunable by the dimensionality of the system.  In revealing superconductivity in a new layered nickelate compound, we unlock the rare-earth nickelates as a family of superconductors beyond the doped infinite-layer compound. While the nickelates are their own class of superconductors distinct from the cuprates, it is intriguing that the cuprate-motivated predictions of optimal $d^{8.8}$ filling have yielded superconductivity, whether achieved through chemical doping or layering dimensionality.  Our work opens up future avenues in which chemical doping and artificial layering may be harnessed in concert to map out and optimise superconductivity.  In this vein, we suggest that exploring the phase diagram around optimal $d^{8.8}$ filling by using Sr$^{2+}$ or Ce$^{4+}$ to chemically dope a layered nickelate may be an exciting route to further expand the family of nickelate superconductors.

\newpage

\bfl {\bf Acknowledgements} \efl
\noindent We thank M. R. Norman and M. Mitrano for discussions.  We also thank K. Lee and H. Y. Hwang for discussions and technical guidance in reduction experiments; D. Erdosy, J. Lee, N. Pappas, S. Thapa and M. Wenny for continued support in reductions; H. Hijazi at the Rutgers University Laboratory of Surface Modification for assistance in Rutherford backscattering spectrometry; and J. MacArthur for electronics support.  This research is funded in part by the Gordon and Betty Moore Foundation’s EPiQS Initiative, Grant GBMF6760 to J.A.Mu. Materials growth was supported in part by the Platform for the Accelerated Realization, Analysis, and Discovery of Interface Materials (PARADIM) under NSF Cooperative Agreement No. DMR-2039380.  Electron microscopy made use of the Cornell Center for Materials Research (CCMR) Shared Facilities, which are supported through the NSF MRSEC Program (No. DMR-1719875). The Thermo Fisher Spectra 300 X-CFEG was acquired with support from PARADIM, an NSF MIP (DMR-2039380), and Cornell University. Nanofabrication work was performed in part at Harvard University's Center for Nanoscale Systems (CNS), a member of the National Nanotechnology Coordinated Infrastructure Network (NNCI), supported by the National Science Foundation under NSF Grant No. 1541959, and in part at the University of Michigan Lurie Nanofabrication Facility.  G.A.P. acknowledges support from the Paul \& Daisy Soros Fellowship for New Americans and from the NSF Graduate Research Fellowship Grant No. DGE-1745303. G.A.P. and D.F.S. acknowledge support from US Department of Energy, Office of Basic Energy Sciences, Division of Materials Sciences and Engineering, under Award No. DE-SC0021925.  Q.S., S.D. and D.C.C. were supported by the STC Center for Integrated Quantum Materials, NSF Grant No. DMR-1231319.  B.H.G, H.P. and L.F.K. acknowledge support by PARADIM, NSF No. DMR-2039380.  A.T.P. acknowledges support from the Department of Defense through the National Defense Science and Engineering Graduate Fellowship (NDSEG) Program.  J.A.Ma acknowledges support from the Arnold and Mabel Beckman Foundation through a Beckman Young Investigator grant.  O.E. acknowledges support from NSF Grant No. DMR-1904716.  A.S.B. and H. L. acknowledge NSF Grant No. DMR-2045826 and the ASU Research Computing Center for HPC resources.  J.A.Mu acknowledges support from the Packard Foundation. 

\bfl {\bf Author Contributions} \efl
\noindent G.A.P., Q.S., C.M.B. and J.A.Mu synthesised the thin-films with assistance from H.P.  G.A.P., D.F.S. and S.D. conducted the reductions with guidance from J.A.Ma.  B.H.G. and L.F.K. characterised the samples with scanning transmission electron microscopy.  G.A.P., A.T.P. and A.Y. performed transport measurements using fabrication assistance from S.N. and J.T.H.  G.A.P., D.F.S., Q.S., D.C.C., A.T.N. and P.S. performed x-ray absorption spectroscopy.  H.L. and A.S.B. performed density-functional theory calculations.  E.M.N., O.E. and A.S.B. constructed the $t-J$ model.  A.S.B. and J.A.Mu. conceived and guided the study.  G.A.P., A.S.B. and J.A.Mu. wrote the manuscript with discussion and contributions from all authors.

\bfl {\bf Competing Interests} \efl
\noindent The authors declare no competing interests.
\clearpage
\newpage

\newpage

\clearpage
\newpage 

\bfl {\bf References} \efl

\newpage
\bfl {\bf \large Methods} \efl

\bfl {\bf Synthesis of layered nickelate thin films} \efl
The elemental fluxes from the effusion cells were first estimated by synthesizing thin films of NiO on MgO and Nd$_2$O$_3$ on yttria-stabilised zirconia.  The thickness of these films was extracted from x-ray reflectivity fits. These deposition rates were also verified using Rutherford backscattering spectrometry.  Using these flux estimates, we then synthesized NdNiO$_3$ on LaAlO$_3$ to refine the stoichiometry \cite{li2021impact}. X-ray reflectivity fits of the optimized NdNiO$_3$ films were used to determine the overall monolayer dosage.  Using these optimized conditions, all layered Ruddlesden-Popper nickelate films were then synthesised by sequentially depositing the neodymium and nickel sources using shutter times found from NdNiO$_3$, starting with neodymium at the interface.  Reflection high-energy electron diffraction (RHEED) was used \textit{in-situ} to monitor the formation of secondary phases and adjust the substrate temperature and/or shutter times of the neodymium and nickel sources accordingly.  Distilled ozone flow (Heeg Vacuum Engineering) was controlled with a piezoelectric leak valve to maintain total chamber pressures in excess of 1.5 x 10$^{-6}$ Torr to ensure the full oxidation of nickel into the 2.67+ or 2.8+ states for the Nd$_4$Ni$_3$O$_{10}$ and Nd$_6$Ni$_5$O$_{16}$ compounds, respectively.  Films were synthesised to contain a total of 60 Ni-O layers, $\sim$27 nm for Nd$_4$Ni$_3$O$_{10}$ and $\sim$25 nm for Nd$_6$Ni$_5$O$_{16}$.  All films were cooled in 1 x 10$^{-6}$ Torr of distilled ozone to discourage the formation of oxygen vacancies.  None of the films were deposited with capping layers as has been performed elsewhere \cite{Li2019,hwang_synthesis,li2021impact}.  We note that the successful synthesis of the Nd$_6$Ni$_5$O$_{16}$ yielding superconducting films is the most sensitive aspect of this experiment, and provide additional details in Supplementary Note 2.

The oxidised Ruddlesden-Popper films were then reduced into the layered square-planar phase using topotactic CaH$_2$ methods similar to those described elsewhere (refs. \onlinecite{Li2019,hwang_synthesis}).  The as-synthesised films were cut into $\sim$2.5 x 5 mm$^2$ pieces; gently wrapped in aluminium foil suitable for ultra-high vacuum (All-Foils Inc); and inserted in borosilicate tubes (Chemglass Life Sciences), pre-baked at 200 \textdegree C, with $\sim$100 mg of CaH$_2$ pellets ($>$ 92\%, Alfa Aesar).  The glass tubes were sealed while being pumped on with a small turbomolecular pump down to $<$0.5 mTorr.  Reductions were conducted in a convection oven (Heratherm, Thermo Fisher Scientific) at 260 - 300 \textdegree C for 3-6 hours with heating rates of 10\textdegree C/min.  After reduction, the films were briefly sonicated in 2-butanone and isopropanol to remove CaH$_2$ residue.  Additional details on the synthesis can be found in Supplementary Note 2.

\bfl {\bf Structural characterisation} \efl
X-ray diffraction (XRD) was performed with a Malvern Panalytical Empyrean diffractometer using Cu K$\alpha_1$ radiation.  Reciprocal space maps (Fig. S3) were taken using a 2D pixel detector (PIXcel3D).  Cross-sectional TEM specimens were prepared using a Thermo Helios G4 UX and/or FEI Helios 660 focused ion beam (FIB) with a final thinning step at 2 keV.  HAADF and ABF-STEM was performed using an aberration-corrected Thermo Fisher Scientific Spectra 300 X-FEG operated at 300 kV with a probe convergence semi-angle of 30 mrad. EELS measurements were carried out at 120 kV on the same instrument equipped with a Gatan Continuum spectrometer and camera.  Any cleaning procedures of the FIB lamellas involving oxygen plasma were foregone for the reduced square-planar samples to prevent reoxidation into the Ruddlesden-Popper phase.  Additional details may be found in Supplementary Note 3.

\bfl {\bf Transport measurements} \efl
Devices in Hall bar geometries were patterned using typical UV photolithography processes. Designs were written on 4” fused silica mask plates using a Heidelberg DWL-2000 mask maker with a 4 mm write head. The photolithography for the ion milling and deposition patterns was done using a conventional SPR 220 based process. All samples were ion milled using an Intivac Nanoquest II Ar ion mill at a working temperature of 15\textdegree C (to prevent sample reoxidation) and an incident beam angle of 10\textdegree. The Ti(10)/Pt(100) (units in nm) metal contact layer was deposited on top of a deposition lithography mask via pulsed laser deposition (PLD) for a standard metal lift-off process.  In some cases, Cr(5)/Au(100) (units in nm) metal contact layers were deposited with an electron-beam evaporator and Hall bar channels defined by a diamond scribe.
Transport measurements down to 1.8 K were conducted in a Quantum Design Physical Property Measurement System (PPMS) equipped with a 9 T magnet using standard AC lock-in techniques at $\sim$15 Hz.  Below 1.8 K, measurements were performed in a Leiden Cryogenics CF-700 dilution refrigerator down to $\sim$50 mK.  To avoid capacitive losses from high excitation AC frequencies exacerbated at low temperatures, resistances below 1.8 K were extracted from current-voltage (I-V) sweeps performed using DC techniques.  4-5 I-V sweeps at each point were averaged to improve the DC signal-to-noise and resistances were taken from the slope fitted across $\pm$ 0.5 $\mu$A.  Hall coefficients were calculated using field sweeps up to 9 T; a representative example is shown in Fig. S10.

\bfl {\bf X-ray absorption spectroscopy} \efl
X-ray absorption spectroscopy (XAS) was performed at 78 K in the total electron yield (TEY) at Beamlines 4.0.2 and 6.3.1 at the Advanced Light Source at Lawrence Berkeley National Laboratory.  All spectra presented are polarisation-averaged and normalised to the signal impinging on a copper mesh.  The spectra are further scaled at the pre-edge and normalised across the entire edge to compare relative intensities.  Each spectrum presented in Fig. S11 represents the average of 8-12 individual scans.

\bfl {\bf Computational Methods} \efl
To calculate the paramagnetic electronic structure of the $n=3$ and $n=5$ layered nickelates, we employ density functional theory (DFT) calculations using the projector augmented wave (PAW) method, as implemented in the Vienna {\it ab initio} Simulation Package (VASP) \cite{VASP}.  We use a pseudopotential which treats the Nd-$4f$ electrons as core electrons. The in-plane lattice parameters were set to match the NdGaO$_{3}$ substrate, and the out-of-plane lattice parameter was optimised. We find that our optimisations agree well with the experimentally measured lattice parameters (see Supplementary Note 6 for more details as well as for the calculations of the Hall coefficient). 

To study the dominant pairing instability for the $n=5$ compound, we solve an effective $t-J$ model for the five NiO$_{2}$ layers using a slave-boson representation and generalizing the spin symmetry of the model to Sp(N) for large N to obtain the superconducting gap order parameter as a function of filling.  We decouple the exchange interactions in both Hartree and pairing channels, and obtain saddle-point solutions as functions of the total nickel doping. A detailed description of the $t-J$ model and its corresponding solutions are presented in Supplementary Note 7.

\bfl {\bf Data Availability} \efl
\noindent The data supporting the findings of this study are available from the corresponding authors upon reasonable request. Source data for Figs. \ref{fig:mainstructural} - \ref{fig:ES} are provided with this paper.

\end{document}